\documentclass[letterpaper,twocolumn,showpacs]{revtex4}
\usepackage{graphicx,bm}

\usepackage{color}
\begin{document}

\title{The spontaneous emergence of ordered phases in crumpled sheets}
\author{Yen-Chih Lin$^{1}$, Ji-Ming Sun$^{1}$, Jen-Hao Hsiao$^{1}$, Yeukuang Hwu$^{2}$, C. L. Wang$^2$, Tzay-Ming Hong$^{1}$}
\affiliation{
  $^{1}$Department of Physics, National Tsing Hua University, Hsinchu, Taiwan 300, Republic of China\\
  $^{2}$Institute of Physics, Academia Sinica, Taipei, Taiwan 115, Republic of China
  }
\date{\today}
\begin{abstract}
X-ray tomography is performed to acquire 3D
images of crumpled aluminum foils. We develop an algorithm to trace out the labyrinthian
paths in the three perpendicular cross sections of the data
matrices. The tangent-tangent
correlation function along each path is found to decay exponentially with an effective persistence length that shortens as the crumpled ball becomes more compact. In the mean time, we observed ordered domains near the crust, similar to the lamellae phase mixed by the amorphous portion in lyotropic liquid crystals.
The size and density of these domains grow with further compaction, and their orientation favors either perpendicular or parallel to the radial direction. Ordering is also identified near the core  with an arbitrary orientation, exemplary of the spontaneous symmetry breaking.
\end{abstract}
 \maketitle
%----------------Introduction--------------------------------------------------
Crumpling is capable of producing a highly rigid
structure with a record minimum of material. Every child knows how to make a
baseball out of a crumpled newspaper. Even with the assistance of a
dutiful parent, its interior shall remain roughly
80\% vacant\cite{Neil1}. In addition to this application, equally fascinating and puzzling phenomena related to crumpling abound in a wide range of length scales, for instance, how DNA is packed in the tiny capsule of
viruses\cite{DNA} or the practical challenge for the
auto industry to come up with a better design to protect the safety of drivers during car accidents\cite{car_crash}.

The enormous resistance of a crumpled ball can be attributed to
the geometrical constraint and the self-avoidance. The
former refers to the inevitable development of D-cones due to the
unstretchability of a thin sheet\cite{Witten07, single}. However, beyond the geometric and
mechanical properties of a single and two-cone interaction\cite{double},
collective behavior of the microstructures such like ridges and
vertices remains unexplored.
% This is why an assessable information on the packing configuration of a self-avoiding sheet will be very useful. 
Previous simulations\cite{Gompper} have identified the phantom and self-avoiding sheets as belonging to different universality classes because the universal exponent $\alpha$ of their force-size relation is differernt. To
clarify the effect of self-avoidance, x-ray tomography
becomes highly desired because it enables us to study
the evolution of the internal structure systematically and perform calculations with its data\cite{Cream_layer,tomography}.

It is known\cite{Witten07} that the macroscopic properties are
shaped more by the collective behavior rather than individual ridges
and vertices. This is because these singularities are linked by not just the tensile force and bending rigidity, but also correlated by the strong hard core interaction\cite{Gompper,Timonen}. When a flat sheet is subject to a gentle force, the
first deformation due to the  buckling is of a conical shape\cite{Timonen2}.
%As we confine the cone into a smaller size, it starts to crumple.
As the compaction progresses, the single cone deformation is followed by a large number of ridges and vertices\cite{Neil1}, while the facets they encircle also
begin to align. 
%The latter trend is reminiscent
%of the lamella phase, except that the ordering is along
%the radial direction. This prompts the formation of domains and is
%thus more appropriate to be categorized as the lamena phase.
 % Although the second response takes no energy, a crumpled structure never has a global ordering, because it minimizes the system's entropy and is occasionally prohibited by the self-avoidance.
In this Letter, we present the first systematic analysis of the local and global structural ordering inside the crumpled aluminum ball. An algorithm is developed to trace out the curves in x-ray tomography
and reconstruct the cross-section view without destroying the sample.
Through the buckling and ordering, we study how they accumulate and affect the final configuration in this highly non-Markov process.

%---------------sample fabrication-----------------------------------------------
Nine sheets of aluminum foil with different diameter ($R_{0} [mm]$= 3, 4,
5, 6, 6.5, 7, 8, 9, 10) are randomly folded by hand first and then squeezed by the flat tip tweezer at different directions\cite{Balankin1} into balls of the
same final radius $R=$1.5 $mm$. 
%Samples of this small size are difficult to fabricate by the high-pressure chamber\cite{Neil1}. 
To determine whether they
still belong to the thin sheet regime, we calculate their Foppl-von
Karman number\cite{Witten07}, $(R_{0}^{2}/h^{2})[12\times(1- \nu^{2})]$ where
$\nu=0.35$ is the Poisson ratio and $h=16\mu m$ is the thickness of
the foil. Ranging from $2.4\times10^{4}$ to $2.6\times10^{5}$, they
turn out to be of the same order as in previous work\cite{Gompper,Timonen,Timonen2,Neil1}.
%Although the diameters of aluminum foil we used are small, they can still be considered as thin sheets because most work done to produce deformation is allocated into the bending energy term.

%---------------X-ray tomography-----------------------------------------------

A special version of microtomography is employed, based on the high intensity x-ray from synchrotron\cite{hwu1}. It provides a standard resolution between 1-2 $\mu$m which shows clear reconstructed images for our analysis. The experiment is performed at the 01A beamline of National Synchrotron Radiation Research Center in Taiwan.  The beamline provides unmonochromatic x-rays  whose energy distribution is 8-15 keV. Image acquisition time per projection is about 10ms, which is captured by a CCD with 2X optical lens focused on a CdWO$_{4}$ single crystal scintillator. The resulted reconstruction consists of a data matrix of 1200$\times$1200$\times$1200 pixels of size 3 $\mu$m.%  Higher resolution image can be obtained by transmission X-ray microscopy which has recently achieved a resolution of 30nm\cite{hwu2}. However, the high resolution comes with a trade-off with reducing specimen size.

%---------------Tracing Algorithm-----------------------------------------------

To study the packing configuration quantitatively, it is necessary to vectorize the
data points. This segmentation method is complex and case-dependent\cite{path}.
The most challenging part in tracing the crumpled surface is to
distinguish two contact planes.
According to previous work\cite{Eric} which concluded that
cross sections through different angles share the same
statistics, we are assured that each of these cross sections is representative of
the  bulk configuration. Therefore, we can focus on the
development of 2D tracing algorithm and present its result 
as a precursor for a full 3D construction from these images.

\begin{figure}[!htb]
\begin{center}
\includegraphics[width=8cm]{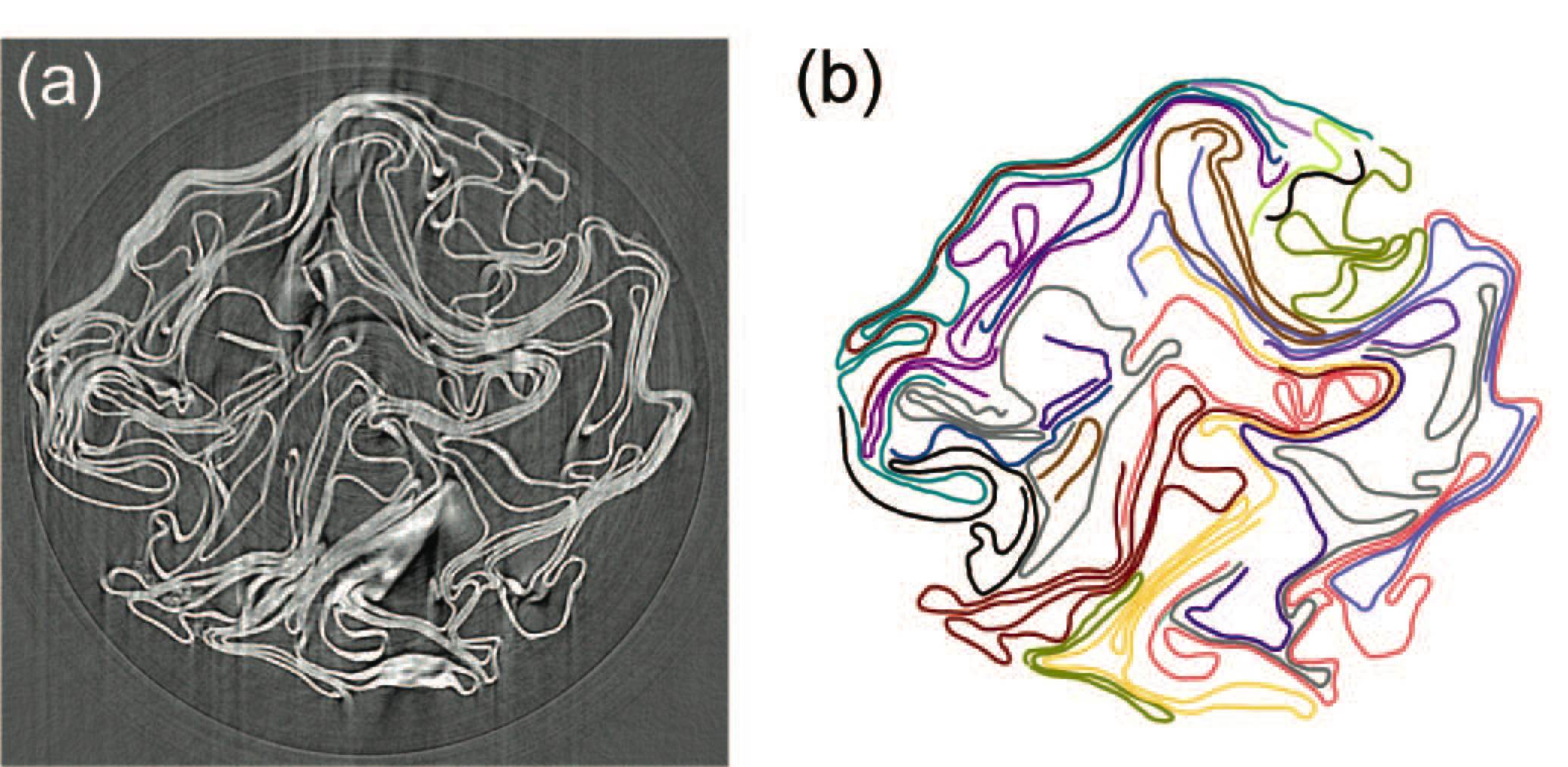}
\caption{Panel (a) shows a slice of raw images reconstructed from 1000 projections for $R/R_0 =0.167$. The glisterns and rings  are experimental artifacts\cite{tomography} that can be reduced by the fill tracing method.
As a contrast, the segmented configuration is shown on panel (b) by linking the traced points.
%The persistence length was defined by calculating the tangent-tangent correlation along the trajectory. The exponential decay implies there is randomization which will wash out the correlation. As the compaction decreasing, the persistence length decreased.
} \label{trace_demo}
\end{center}
\end{figure}

 After all the images have been reconstructed into
3D data matrices, we resample three perpendicular
cross sections which are X-Y, Y-Z, Z-X planes of the sample.
Briefly, the procedures of tracing algorithm are mainly divided
into three parts: random seeding, identifying solutions, and
labeling traced points. We start with a circle around the
seed point with an radius of 24 pixels, and then find all crossing points
between its perimeter and
the paths. Since many paths can be close to each other, multiple solutions frequently occur. When this happens,
we select the point that is joined to the seed point from all
candidate solutions. The traced points are marked immediately to
avoid double tracing with the forest-fire simulation. After all the
points have been vectorized, we perform a high order Beizer fitting
to resample the segmented points to reduce irregularities and
increase the sampling rate, see Fig.\ref{trace_demo}. A more detailed description of the
tracing algorithm and segmented images can be found in the
online supplemental document\cite{path}.

\begin{figure}[!htb]
\begin{center}
\includegraphics[width=7cm]{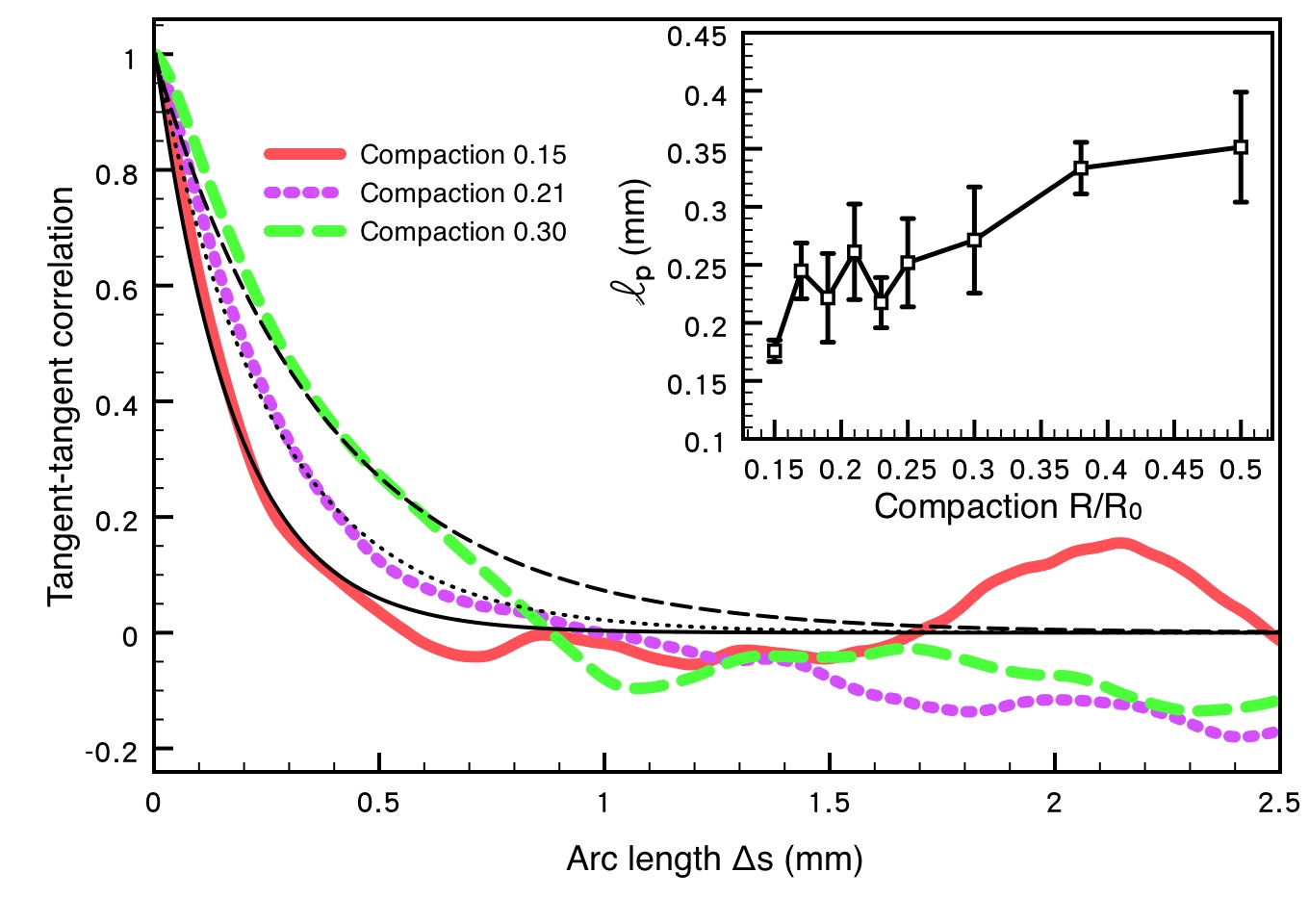}
\caption{ Tangent-tangent correlations $C(\Delta s)$ are plotted for
three different compactions $R/R_{0}$. Although the data fluctuates in the tail, which is likely an artifact of too few samplings of long paths, an exponential decay and a
persistence length can be extracted. Thick lines illustrate the experimental data, and
thin lines are the fitting curves. 
%The persistence length was defined by calculating the tangent-tangent correlation along the trajectory.
%The exponential decay results from the randomization which tends to wash out the correlation. 
The inset highlights the decrease of
persistence length as the ball shrinks.  } \label{persistence length}
\end{center}
\end{figure}

We start by calculating the tangent-tangent correlation function which is a basic statistical
property related to the buckling\cite{semi} of sheets. This function is defined
as $C(s,s^{\prime})\equiv \langle(\mathbf{u}(s)-\langle
\mathbf{u}(s)\rangle)\cdot (\mathbf{u}(s^{\prime})-\langle
\mathbf{u}(s^{\prime})\rangle ) \rangle$, where
$\mathbf{u}(s)=d \mathbf{R}/ds$ is the tangent
vector at arc length $s$ in the curvilinear coordinate and
$\mathbf{R}(s)$ is the position vector. It can
be simplified to $C(|s-s^{\prime}|)=C(\Delta s)$ if the system exhibits translational invariance. 
%We measure this correlation as the form $C(\Delta s)$ to obtain more data for ensemble. 
This assumption was checked quantitatively to hold
except at the end points where the fluctuations become large.

The function $C(\Delta s)$ is found to decay exponentially in Fig.\ref{persistence length} with an effective persistence length
$l_{p}$. 
%Unlike the ground state in crumpling, $C(\Delta s)$ decays exponentially without any sinusoidal oscillation. The decay 
This decay form can be derived from the random packing of facets.
%paths decreases with the increasing of path length, the size of
%ensemble sample is less for larger $\Delta s$. This  causes
%$C(\Delta s)$ to fluctuate in large distance $\Delta s$.
%(Ren-Hao will add the derivations we did this afternoon in the department office)
Borrowing the concept of tube model for polymers\cite{Gennes}, one can think of the sheet as moving inside two walls which model the confinement due to the hard core interaction from its neighboring portions of sheet. When we cut perpendicularly through the walls, the cross section will reveal a wiggling path with a static configuration similar to that of a polymer in the tube model. However, it should be noted that their dynamics are different because the path is, afterall, a projection of a 2D sheet. 
The movement of each segment needs to coordinate with the rest of the sheet, unlike the reptation model in which the polymer is confined in all sidewise directions by a static tube.
The spacer width apparently decreases as we increase the crumpling force. Dividing this width by the segment length $a$ of the path gives the maximum angle $\theta$ between neighboring segments. Roughly, we can imagine the configuration as being mapped out by a random walk with a fixed stride $a$ but only two choices of angular deviations, $\pm \theta$. Then the probability of finding the relative angle between the $n$th and zeroth segments equaling $m\theta$ obeys the Gaussian distribution for a random walk after $n\equiv (s^\prime -s)/a$ steps: $P(m,n)\simeq\exp(-m^2/2n)$. The function
$C(\Delta s)=\langle\cos(m\theta)\rangle$ can be explicitly evaluated:
%\equiv \langle\mathbf{u}(0)\cdot\mathbf{u}(\Delta s=na)\rangle\langle 
\begin{equation}
\sqrt{{2\over \pi n}}\int\cos(m\theta)\exp\Big({-{m^2\over 2n}}\Big)dm=\exp\Big(-{na\over l_{p}}\Big)\nonumber
\end{equation}
where $l_{p}=2a/\theta^{2}$.  The
inset of Fig.\ref{persistence length} shows that $l_{p}$ shortens
as $R/R_0$ decreases, which implies the crumpled structure
becomes more disordered  along the curvilinear direction. This observation also requires that the segment length $a$ not only depends on the bending rigidity of the material, but can be cut short by the compact packing. 
%suffers from larger deformation more frequently when the sample is more compact, and the correlation drops faster. 

 The concept of an effective persistence length is similar to that in  polymers\cite{Gennes} and  the de Gennes coherence length in membranes\cite{coherence}.
%conventioanl persistence length in polymer and in membrane,
However, different from the latter two cases, our crumpled ball is so macroscopic\cite{Edwards theory} that the thermal temperature becomes irrelevant. Instead, it is the noise from the random folding that allows the sheet to appear rumpled. This zero-temperature randomizing effect also exists in the granular systems\cite{Edwards}. 
% because crumpling is intrinsically a series of random processes starting from the uncertainty in the sample preparation, not to mention possible inhomogeneities on the sheet.

\begin{figure}[!htb]
\begin{center}
\includegraphics[width=5cm]{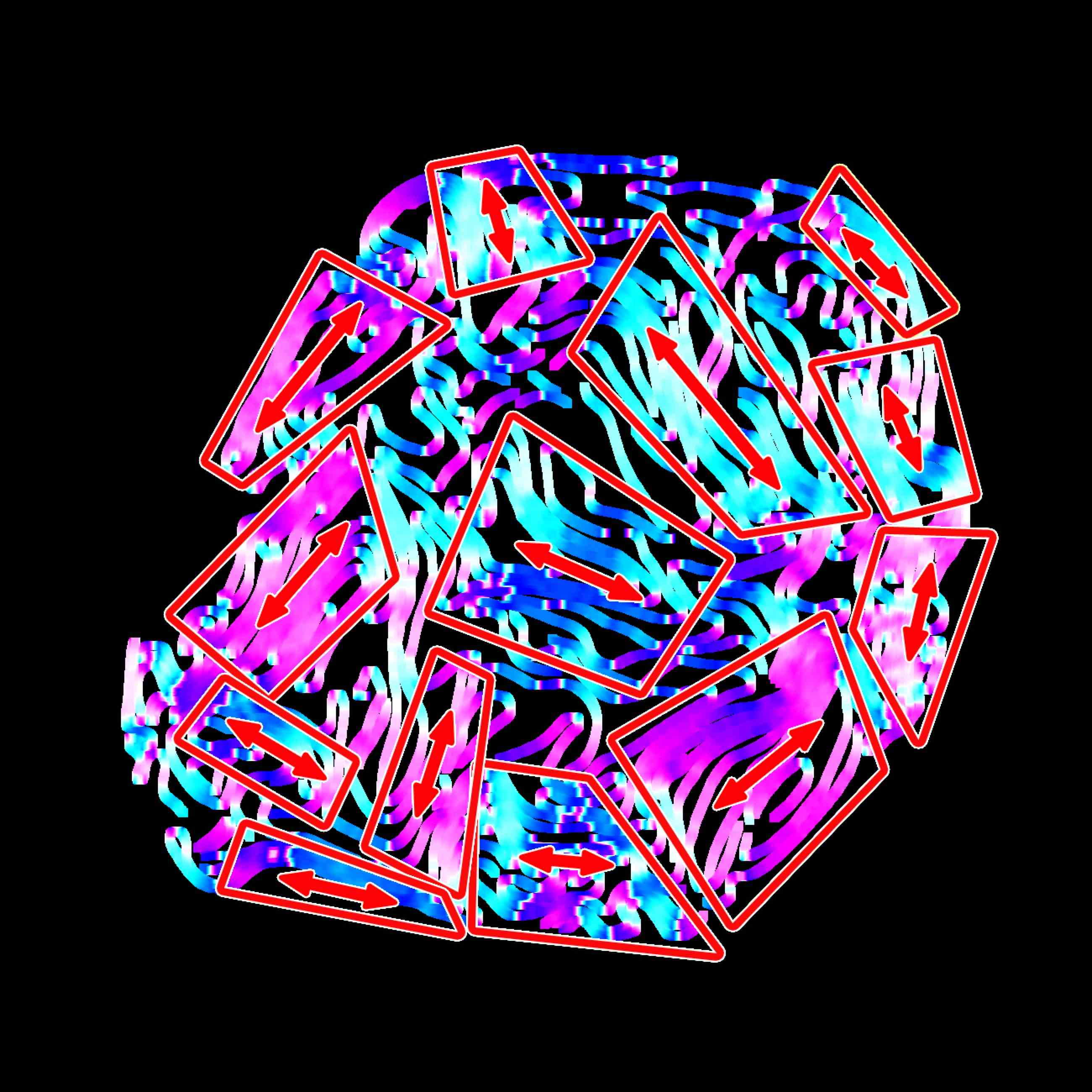}
\caption{We trace out trajectories and define different domains marked by quadrangles for $R/R_0 =0.15$. Different color or graylevel signalizes different orientations. The domain formation caused by the hard core interaction and high density of paths is similar to that of lyotropic liquid crystals\cite{lyotropic}.} \label{Domain}
\end{center}
\end{figure}

\begin{figure}[!htb]
\begin{center}
\includegraphics[width=8cm]{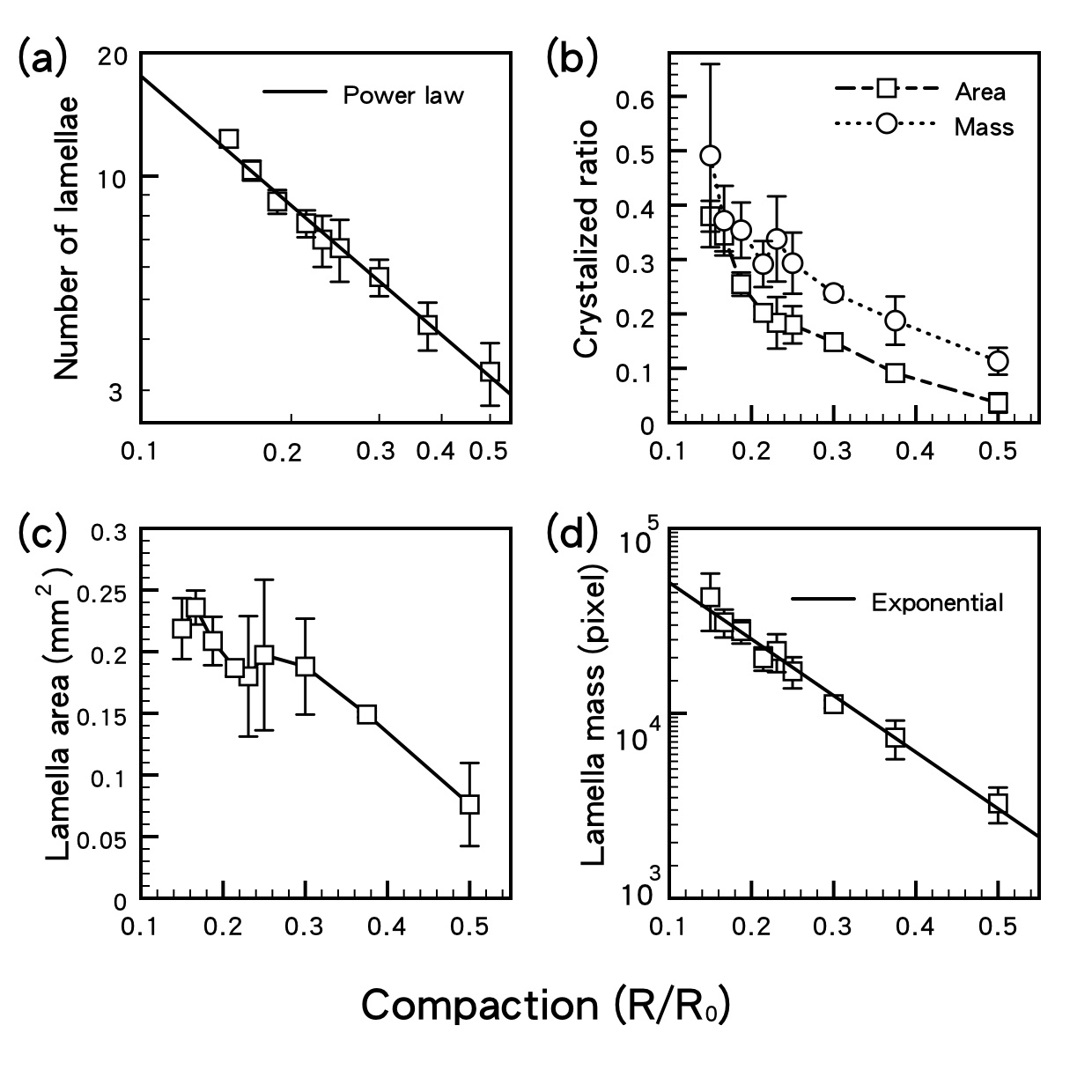}
\caption{Four properties of the lamella-like phase are calculated as a function of $R/R_0$. They include the number of lamellae in panel (a), ratio of total area and mass covered by the lamellae in (b), size of each lamella in (c), and  mass  or total length encompassed in each lamella in (d). Data in (a) and (d) can be fit by $1.57 (R/R_{0})^{-1.05}$ and $10^5 \exp (-7.04R/R_0 )$.} \label{lamella}
\end{center}
\end{figure}

After calculating the order along the curvilinear coordinate, we
turn to the cartesian plane. Note that
locally ordered structures can be identified in Fig.\ref{trace_demo}. The paths in the plane, which cut through
the facets in the 3-D sample, show a tendency to align and form
lamella-like phase. We can separate the ordered portion from the disordered one to define crystalline and amorphous regions.
This classification is aided by the vectorized data. In
Fig.\ref{Domain}, all paths are denoted by different colors and
brightnesses in grayscale to indicate different orientation
to the $(1,0)$  direction. The neighboring facets with
the same orientation are marked with the same color, and can be easily
recognized as a lamella. %On the other hand, the paths in theamorphous part are shaded with random color.
 To extract the domain
boundaries, the Laplacian of the color
brightness was first calculated with the local maxima signalizing
the location of the divisions. Although the distribution of lamellae can in principle be derived by this method, manual identifications are still required when the color gradient is too noisy. Therefore, we use the quadrangles to label all
lamellae with the reference of Laplacian field to obtain the data
with a coherent format.

Using the data averaged
over three perpendicular cross sections for each sample, we calculate four essential properties of the lamellae: their number, ratio of total area and mass they covered in the cross section, their size, and the mass encompassed in each lamella. We checked that the product of data in panels (a) and (c) equaled the area ratio  in (b) times the cross section area, $\pi R^2$.
Figure \ref{lamella} shows that they all increase as $R/R_0$ decreases. 
Panel (a) indicates the number of lamellae is inversely proportional to $R/R_0$. Two features are worth noting in (b): Firstly, the reason why the two lines are not plainly proportional is that the total mass grows as we fix $R$ and increase $R_0$ to achieve lower ${R/R_0}$. In contrast, total area $\pi R^2$ is unchanged. Furthermore, the alignment allows for more efficient packing inside the domain which explains why the mass ratio is larger. Secondly, both ratios never exceed 0.5. This is similar to supercooled liquid where the extent of ordering is hindered from being complete\cite{glass}.
One may wonder how a structure with so much amorphous region can be so hard. A possible explanation is that these ordered domains near the boundary interlock and act like a hard crust.  Each domain consists of many aligned layers which greatly enhance the bending rigidity and make them more resistant to buckling.

\begin{figure}[!htb]
\begin{center}
\includegraphics[width=6.5cm]{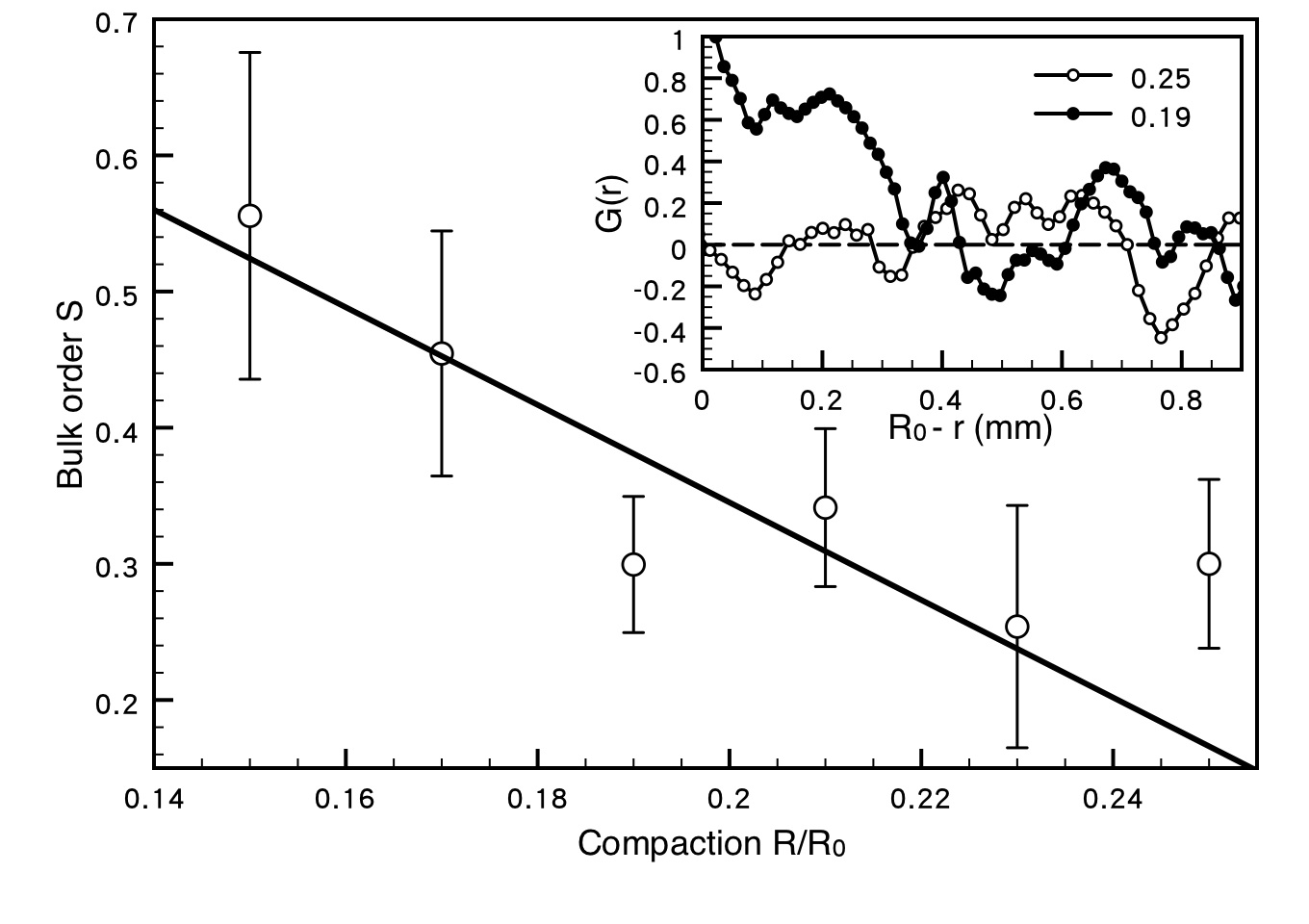}
\caption{The bulk order $S$ is shown to grow when the sample becomes more compact. The solid line is a guide to the eyes to emphasize the trend that the ordering increases as the ball becomes more compact. 
The $R_0$(mm)=3, 4, and 5 data are not inlcuded because they no longer exhibit a shell region and also their error bars are too large due to insufficient samplings. 
%Spurious contributions to $S$ from the finite number effect\cite{Jennifer} are estimated to be less than $10^{-5}$ and so are insignificant.
%Both effects make it hard to determine the nature of this transition.
The inset shows the wall-rod correlation functions $G(r)$ for $R/R_0=$0.25 and 0.19. 
%We find that the higher the compaction (more loose packing), the larger fluctuation without a decay trend is observed. 
%Another reason causes the fluctuation is that less mass are averaged for a loose packing sample. 
%The region with $G(r)>0$ is defined as the shell, while the negative one the bulk\cite{Jennifer}. 
% This is in contrast to the shell region which grows with further compression and where the lamellae mostly reside. 
%If no sudden drop can be detected in $G(r)$, it means the crust of the sample is really thin and the whole sample can be regarded as bulk.
} \label{order}
\end{center}
\end{figure}

In the previous work\cite{Jennifer} that demonstrates a spontaneous patterning in vibrated rods, a wall-rod correlation function $G(r)\equiv \langle cos(2 \eta) \rangle$ was defined to extract the size of the core or bulk area, where $\langle \rangle$ denotes averaging over $\varphi$ and $\eta$ is the angle between the tangent vectors at point ($r,\varphi$) and the boundary in the $\varphi+\pi/2$ direction. We repeat the same definition to study the packing configuration near the core and plot the results in the inset of Fig.\ref{order}. Since the external force acts from outside, it is natural that the correlation with the boundary decreases as we enter the core. The division between the bulk and shell regions is marked by the first vanishing of $G(r)$. 
%how deep in the sample that remains correlated with the boundary, see Fig.\ref{order}.
%For instance, Nearby the boundary of a highly compact sample, $G(r)$ is almost unity which means that the sheets almost align to the boundary. 
Again, following the notation of \cite{Jennifer}, an order parameter $S$=$\langle\cos(2\theta)\rangle$ is defined for the bulk where $\theta$ is the angle between the tangent vector and lamella direction of the bulk.
%Although we expect that the structure around the center should be disorder because no effect from the boundary penetrates, an
%ordered configuration at the bulk area, however, 
We are surprised to find a spontaneous bulk ordering in Fig.\ref{order} where correlations with the boundary layer have considerably weakened. As $R/R_{0}$ decreases, the bulk order
 increases. The ordered phase induced by high concentration is similar to that happened in \cite{Jennifer} and lyotropic
liquid crystals. According to Onsager\cite{onsager}, although parallel arrangements of anisotropic objects lead to a decrease in orientation entropy, there is a gain in positional entropy. Thus, a positional order is expected to become entropically favorable at sufficient rod concentrations.

%------------------------Discusssion------------------------------------------

Given that the deformations in aluminum are plastic and irreversible,  crumpling can be viewed as a series of quenching process since not all configurations are accessible, nor equally probable. The noise introduced by the random folding still enables the sheet to slightly adjust its configuration to seek a local potential minimum\cite{ET}. 
%This makes it possible to discuss the probability distribution of the ridge length or facet size in each step of the crumpling process\cite{paper}. 
%Even though a crumpled sheet does not obey ergoticity, the process of random folding enforces the system explore the phase space for possible configurations\cite{ET}.
This random process plays the role of vibration in the spontaneous patterning of vibrated rods\cite{Jennifer} and slow shearing in granular systems\cite{Edwards,Makse}. 
Since this large amount of facets are not only correlated by the sheet but also interact strongly via the bending potential, the geometrical constraint forbids our crumpled sheet from reaching the true ground state\cite{rod}. Consequently, the bulk order parameter of our compact sample is still much lower than that of vibrated rods\cite{Jennifer}. 
%The packing configuration reported in this Letter also suggests that a crumpled structure is quasi-equilibirium. Although more rigorous examinations of this statement, such like calculating the free energy or applying further shearing on the sample, are needed, to our knowledge, this is the first experimental evidence which supports the hypothesis adopted in previous simulation\cite{Eric}. 
%The spontaneous emergence of structural order is discussed to be intimately linked to the strong resistance of the crumpled ball. 
% And the compaction plays the same role as the temperature\cite{Edwards}.

%-------------------------Conclusion------------------------------------------
In conclusion, we performed x-ray tomography to quantitatively
study the inner structure of aluminum foils at different
compactions. All paths in the three perpendicular cross sections are
traced out and vectorized before calculating the statistical
properties of their packing configuration. The tangent-tangent
correlation of the path reveals an effective persistence length that decays with
the compaction. A second length scale associated with
%the crumpled sample becomes more compact, we
the size of domains that emerges near the crust and mimics the lamella phase in
lyotropic liquid crystals. Number of these domains and their area and
mass grow monotonically with the compaction, and their orientation favors either perpendicular or parallel to the radial direction.
We also identified an ordered domain near the core of the crumpled ball with an arbitrary orientation, exemplary of the spontaneous symmetry breaking.

We benefit from fruitful discussions with A. S. Balankin and Peilong Chen. Support by the National Science Council in Taiwan under Grant No. 95-2112-M007-046-MY3 and 98-2112-M007-005-MY3 is acknowledged.
%------------------------Bibliography-------------------------------------------

\end{document}